\documentclass{article} 
\usepackage{iclr2024_conference,times}

\usepackage{amsmath,amsfonts,bm}

\def\eqref#1{equation~\ref{#1}}

\def\1{\bm{1}}

\DeclareMathAlphabet{\mathsfit}{\encodingdefault}{\sfdefault}{m}{sl}
\SetMathAlphabet{\mathsfit}{bold}{\encodingdefault}{\sfdefault}{bx}{n}

\usepackage{hyperref}
\usepackage{url}

\usepackage{graphicx}
\usepackage{booktabs} 

\title{Grapevine Disease Prediction Using Climate Variables from Multi-Sensor Remote Sensing Imagery via a Transformer Model}

\author{%
  Weiying Zhao \\
  Deep Planet \\
  London, UK \\
  \texttt{weiying@deepplanet.ai} \\
  \And
  Natalia Efremova\\
  Queen Mary University and the Alan Turing Institute \\
  London, UK \\
  \texttt{natalia.efremova@gmail.com} \\
}

\iclrfinalcopy 
\begin{document}

\maketitle

\begin{abstract}



Early detection and management of grapevine diseases are important in pursuing sustainable viticulture. This paper introduces a novel framework leveraging the TabPFN model to forecast blockwise grapevine diseases using climate variables from multi-sensor remote sensing imagery. By integrating advanced machine learning techniques with detailed environmental data, our approach significantly enhances the accuracy and efficiency of disease prediction in vineyards. The TabPFN model's experimental evaluations showcase comparable performance to traditional gradient-boosted decision trees, such as XGBoost, CatBoost, and LightGBM. The model's capability to process complex data and provide per-pixel disease-affecting probabilities enables precise, targeted interventions, contributing to more sustainable disease management practices. Our findings underscore the transformative potential of combining Transformer models with remote sensing data in precision agriculture, offering a scalable solution for improving crop health and productivity while reducing environmental impact.

\end{abstract}

\section{Introduction}

Remote sensing technology, with its integration of multisensor image analysis and climate feature assessment, has become a cornerstone of precision agriculture, particularly in viticulture, where it offers unparalleled advantages in the early detection of diseases. By leveraging multispectral  \cite{ferro2023assessment}, thermal imaging technologies \cite{fevgas2023detection}, and climate data, researchers can monitor the subtle spectral and thermal changes in grapevine foliage-early indicators of phytopathological stress. These changes, indicative of disease onset, occur before symptoms are visibly detectable, providing a critical window for early intervention. Multisensor imagery captures the vineyard's detailed spectral profile through various wavelength bands and incorporates climate variables, enriching the analysis with environmental context. This comprehensive approach, when further analyzed with indices such as NDVI and NDWI \cite{zhang2019monitoring}, transforms complex datasets into interpretable metrics closely linked to the vines' health and vitality.

Current disease detection methods in viticulture primarily rely on a combination of manual inspections, laboratory analysis, and remote sensing technologies. Researchers focus on utilizing high spatial or spectral resolution images \cite{ferro2023assessment, kanaley2023assessing} to identify vineyard diseases, with a particular emphasis on detecting individual diseases and analyzing information at the pixel level  \cite{kerkech2020vine}.


This study explores the benefits of utilizing multi-sensor and multi-scale remote sensing features to predict grapevine diseases at the block level. We highlight the significant advancements made possible by these technologies in promoting vineyard health and productivity. A particular challenge in this field has been the lack of extensive in-situ examples for disease forecasting. The introduction of the TabPFN method \cite{hollmann2023tabpfn} marks a substantial improvement in addressing small-scale tabular classification challenges. By integrating the TabPFN model with a block-wise disease database prepared by multi-sensor imagery, we aim to surpass the limitations of current disease detection methods, setting a new standard for accuracy in vineyard disease management.

\section{Methodology}


\begin{figure}[h]
  \centering
  \includegraphics[width=0.85\textwidth]{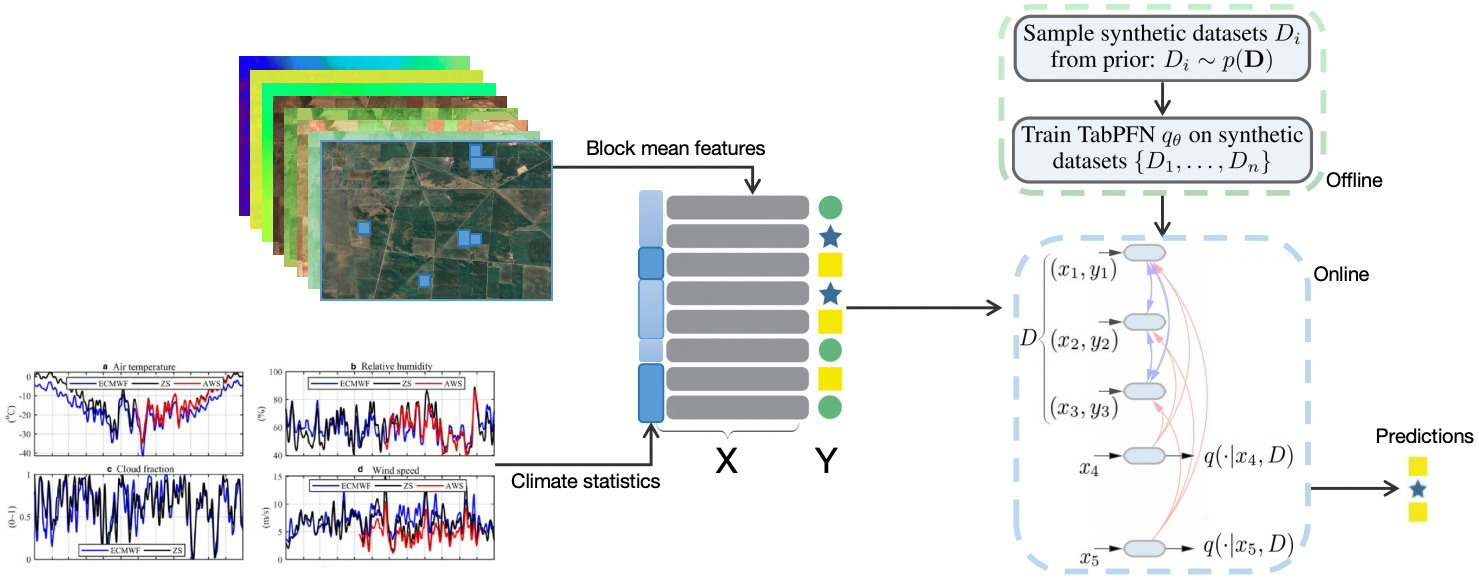}
  \caption{Flowchart of the proposed disease forecasting framework. The TabPFN method \cite{hollmann2023tabpfn,picard2024fast}  is used as an example.  The TabPFN learns to approximate the PPD of a given prior in the offline stage to yield
predictions on a new dataset in a single forward pass in the online stage.}
  \label{fig:TabPFN_flowchart}
\end{figure}


This workflow  (Fig.\ref{fig:TabPFN_flowchart}) illustrates our method progression from initial data preparation to the eventual unknown data classification.  The TabPFN, a Transformer  \cite{vaswani2017attention} based model which contains 12 layers, is designed for classification tasks on small tabular datasets and trained offline. 
It learns to approximate the posterior predictive distribution (PPD) of Bayesian inference on synthetic datasets, which are drawn from a specified prior. In the Bayesian framework for
supervised learning, the prior defines a space of hypotheses on the relationship of a set of inputs
$x$ to the output labels $y$. 
The TabPFN is trained once and can make predictions in less than a second without hyperparameter tuning. It accepts training and test samples as input and produces predictions in a single forward pass, making it competitive with state-of-the-art methods while being significantly faster.

Specifically, given a set of training samples $D_{{train}} := \{(x_1, y_1), \ldots, (x_n, y_n)\}$, the PPD for a test instance $x_{{test}}$ is denoted by $p(y_{{test}} | x_{{test}}, D_{{train}})$.  This PPD is calculated by integrating over the hypothesis space $\Phi$, weighting each hypothesis $\phi \in \Phi$ by its prior probability $p(\phi)$ and the likelihood $p(D|\phi)$ of the data $D$ given $\phi$: \cite{muller2021transformers}:

\begin{equation}
p(y|x, D) \propto \int_{\Phi} p(y|x, \phi)p(D|\phi)p(\phi)d\phi
\end{equation}

During inference, the trained model is applied to unseen real-world datasets. For a novel dataset with training samples $D_{train}$ and test features $x_{test}$, feeding $⟨D_{train}, x_{test}⟩$ as
an input to the model yields the PPD $q_\theta(y|x_{test}, D_{train})$ in a single forward-pass \cite{hollmann2023tabpfn}. The PPD class probabilities are then used to make predictions for the real-world task. To generate synthetic classification labels for imbalanced multi-class datasets, scalar labels \( \hat{y} \) are transformed into discrete class labels \( y \) by dividing the \( \hat{y} \) values into intervals that correspond to class boundaries based on class labels.

\section{Experiments}
\subsection{Dataset}

\begin{figure}[h]
  \centering
  \includegraphics[width=0.7\textwidth]{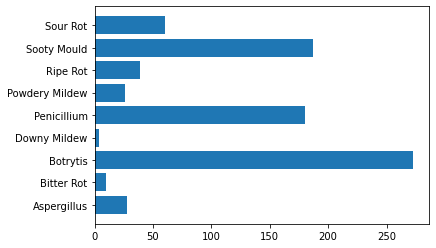}
  \caption{Total blocks affected by different kinds of diseases. The diseases have imbalanced distribution.}
  \label{fig:disease_cases}
\end{figure}

We compared the proposed method with other approaches using a database of diseases measured during two seasons in 76 vineyards in Australia, which contain 627 blocks.  There were around nine different kinds of diseases (Fig.\ref{fig:disease_cases}), such as Aspergillus, Bitter Rot, Botrytis, Downy Mildew, Penicillium, Powdery Mildew, Ripe Rot, Sooty Mould, Sour Rot.  The disease data are measured at the block level. With reference to microbial biogeography for grapevine \cite{liu2019vineyard}, we prepared the following climate-related features with different remote sensing data \cite{zhao2023soil}:
\begin{itemize}
 \item  Spectral features and different vegetation indices are provided by Sentinel-2, which were acquired near the sample time
\item  Climate features provided by ECMWF and MODIS contains macroclimate and microclimate features. Part of them only captures information from the season start time to the disease measuring time.
\item  Soil attributes like Soil type, soil nutrients, soil carbon, pH, bulk density, available water capacity, etc.
\item Terrain attributes like DEM, slope, aspect, etc.
\item Block attributes like variety, row direction, geolocation, etc.
\end{itemize}

The resulting tabular format dataset comprises 1335 samples with around 450 features each. 

\subsection{Experimental results and discussion}
The data were divided into train (76\%) and test (24\%) datasets. The total training data is limited by the TabPFN method \cite{hollmann2023tabpfn}. Since many diseases happened simultaneously, we transferred the target to binary values. The number of disease-affecting blocks is much less than that of healthy blocks. We'll split the method comparison into two groups, with or without balanced or weighted targets.
The top 25 features are selected with SHAP (SHapley Additive exPlanations) values based on the tree-based models. 

TabPFN streamlines input preprocessing by implementing feature-wise standardization, heuristic log-scaling for outliers, and PowerTransform across all attributes for ensemble members. These steps are vital for aligning real-world data with synthetic training datasets. The tree-based model preparation is referenced \cite{mcelfresh2023neural}. We also prepared two popular deep learning methods, Transformer and MLP, for comparison.

\begin{table}[ht]
  \caption{Performance of algorithms over the disease dataset. 
  The assessment includes mean and standard deviation for accuracy, balanced accuracy, and F1-score, computed over 40 iterations with varied seeding for dataset partitioning, to present a comprehensive picture of each model's robustness in balanced and imbalanced classification contexts.}
  \label{tab:spatial_interpolation}
  \centering
  \small
  \begin{tabular}{llllll}
    \toprule
Methods	&Parameters&	Target &Accuracy&	Balanced accuracy&	F1-score\\
\midrule
XGBClassifier&	default&	imbalance&	0.7942±0.0205&	0.7482±0.0266&	0.6538±0.0397\\
XGBClassifier&	default&	balanced&	0.7940±0.0215&	0.7612±0.0254&	0.6707±0.0349\\
LGBMClassifier&	default&	imbalance&	\textbf{0.7972±0.0215}&	0.7505±0.0269&	0.6576±0.0373\\
LGBMClassifier&	default&	balanced&	0.7925±0.0220&	0.7607±0.0287&	0.6693±0.0389\\
CatBoostClassifier&	default&	imbalance&	0.7962±0.0182&	0.7436±0.0219&	0.6482±0.0326\\
CatBoostClassifier&	default&	balanced&	0.7931±0.0233&	\textbf{0.7843±0.0263}&	\textbf{0.6961±0.0358}\\
\midrule
PFNClassifier&	32 ensembles&	imbalance&	\textbf{0.7948±0.0234}&	0.7477±0.0277&	0.6537±0.0398\\
PFNClassifier&	default&	imbalance&	0.7947±0.0243&	\textbf{0.7489±0.0292}&	\textbf{0.6550±0.0416}\\
MLP&	633 trainable&	imbalance&	0.7327±0.0268&	0.7161±0.0294&	0.6063±0.0478\\
MLP&	633 trainable&	balanced&	0.7126±0.0289&	0.7254±0.0292&	0.6187±0.0392\\
Transformer&	669 trainable&	imbalance&	0.6999±0.0254&	0.5946±0.0324&	0.3884±0.0782\\
Transformer&	669 trainable&	balanced&	0.6917±0.0292&	0.6888±0.0340&	0.5756±0.0402\\
    \bottomrule
  \end{tabular}
\end{table}


As shown in Tab.\ref{tab:spatial_interpolation}, the PFNClassifier, when configured with 32 ensembles, shows a competitive accuracy on the imbalanced dataset, with a slight variation in balanced accuracy and F1-score, illustrating the potential of ensemble methods in enhancing prediction reliability. Despite having many trainable parameters, the MLP and Transformer models lag behind the aforementioned algorithms in performance across all metrics. 


Among the algorithms evaluated, the LGBMClassifier with default parameters on the imbalanced dataset achieved the highest accuracy, indicating its robustness in handling imbalanced data without the need for balancing techniques. This is closely followed by the performances of the CatBoostClassifier and the XGBClassifier, both also evaluated under default settings. Notably, the CatBoostClassifier exhibited superior performance on the balanced dataset, achieving the highest balanced accuracy and F1-score, underscoring its effectiveness in leveraging the balanced dataset to improve predictive performance.

\begin{figure}[h]
\graphicspath{{results/}}
   \centering
\begin{tabular}{cccc}
\multicolumn{2}{c}{\includegraphics[width=6.5cm]{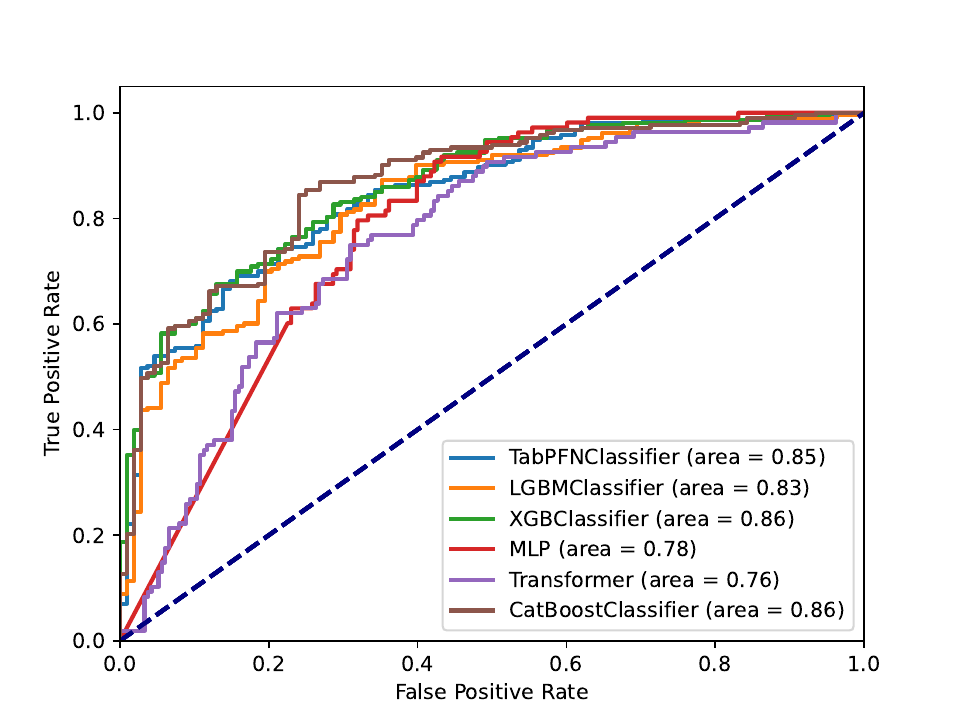}}&
\multicolumn{2}{c}{\includegraphics[width=6.5cm]{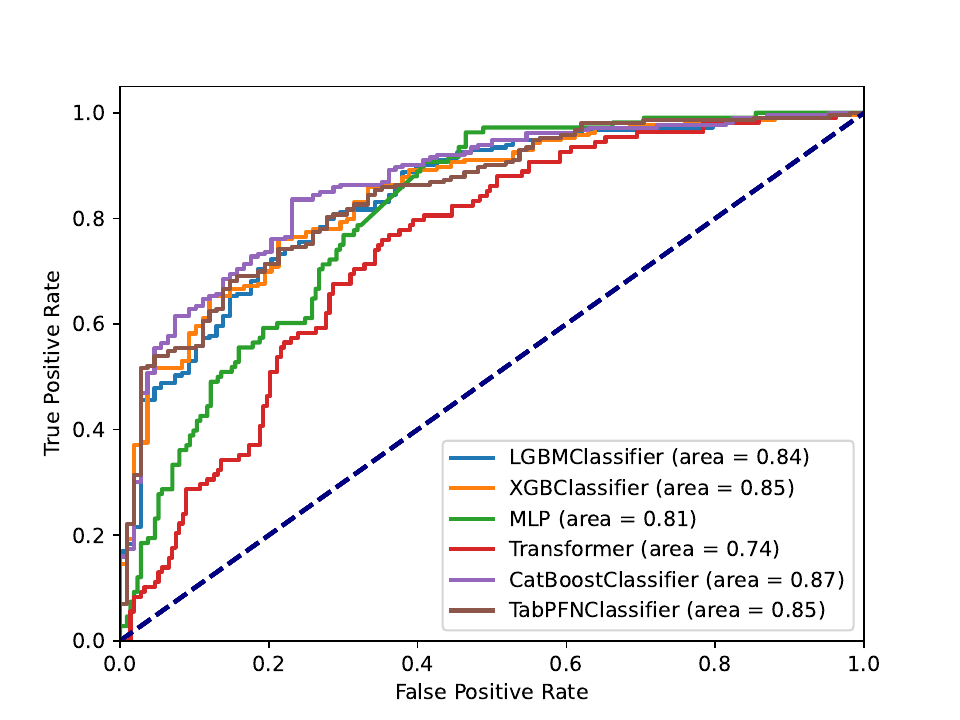}}\\
 \multicolumn{2}{c}{(a) with imbalance data}& \multicolumn{2}{c}{(b) with balanced data}\\
\end{tabular}
  \caption{ROC curve results comparison of the methods. The experimental results are based on the same training and testing databases.}
  \label{fig:training_performance} 
\end{figure}

In the imbalanced dataset scenario, the TabPFNClassifier achieves an AUC of 0.85 (Fig.\ref{fig:training_performance}), showcasing its robustness in dealing with class imbalance. The MLP, however, only attains an AUC of 0.78, indicating potential challenges in handling imbalanced class distributions. For the balanced dataset, the CatBoostClassifier demonstrates superior performance with the highest AUC of 0.87, suggesting an exceptional capability to distinguish between classes. Conversely, the Transformer model shows the least effective performance with an AUC of 0.74, implying room for improvement in its classification power.

These findings suggest advanced models like MLPs and Transformers have shown remarkable success in various domains. However, their application to imbalanced disease datasets without task-specific tuning might not yield optimal results. In contrast, PFNClassifiers demonstrate a superior ability to handle imbalanced data effectively, even without explicit balancing techniques. It can provide comparable results as gradient-boosted
decision trees.

\begin{figure}[h]
  \centering
  \includegraphics[width=1.0\textwidth]{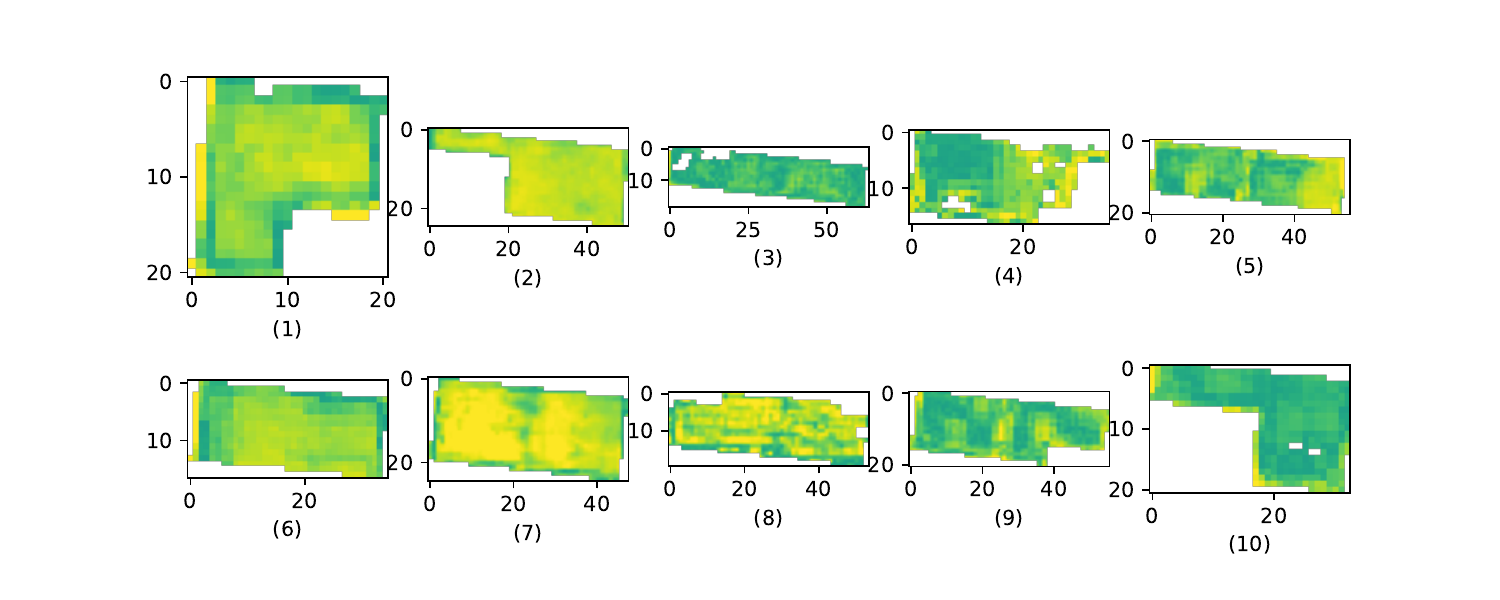}
  \caption{Disease probability maps for 10 blocks in early 2021 in Australia. The probability of disease is displayed as a visual heat map (green = low probability, yellow = high probability).}
  \label{fig:TabPFNClassifier_forecasting_results_03}
\end{figure}



Given the PFNClassifier's capability to output both a binary target and probability estimates, it is suitably applied to detailed, pixel-wise disease forecasting as depicted in Figure \ref{fig:TabPFNClassifier_forecasting_results_03}. This figure presents a collection of pixel-wise classification heatmaps, each corresponding to a specific vineyard block, to visually represent the predicted distribution of disease risk. The gradations of color within these heatmaps delineate the likelihood of disease occurrence, offering a refined, localized risk assessment at the level of individual vineyard blocks.

\section{Conclusion}

This study presents a pioneering approach to predicting grapevine diseases using climate variables from multi-sensor remote sensing imagery, leveraging the TabPFN model. Our findings demonstrate the model's efficacy in processing small, imbalanced datasets, showcasing comparable or superior performance to traditional methods like gradient-boosted decision trees. This approach enhances blockwise disease forecasting in viticulture by incorporating environmental data and advanced machine learning, offering a nuanced understanding of disease dynamics. Future work will pay more attention to using phenology and temporal climate features.

\bibliography{iclr2024_conference}

\begin{thebibliography}{12}
\providecommand{\natexlab}[1]{#1}
\providecommand{\url}[1]{\texttt{#1}}
\expandafter\ifx\csname urlstyle\endcsname\relax
  \providecommand{\doi}[1]{doi: #1}\else
  \providecommand{\doi}{doi: \begingroup \urlstyle{rm}\Url}\fi

\bibitem[Ferro et~al.(2023)Ferro, Catania, Miccich{\`e}, Pisciotta, Vallone, and Orlando]{ferro2023assessment}
Massimo~V Ferro, Pietro Catania, Daniele Miccich{\`e}, Antonino Pisciotta, Mariangela Vallone, and Santo Orlando.
\newblock Assessment of vineyard vigour and yield spatio-temporal variability based on uav high resolution multispectral images.
\newblock \emph{Biosystems Engineering}, 231:\penalty0 36--56, 2023.

\bibitem[Fevgas et~al.(2023)Fevgas, Lagkas, Argyriou, and Sarigiannidis]{fevgas2023detection}
Georgios Fevgas, Thomas Lagkas, Vasileios Argyriou, and Panagiotis Sarigiannidis.
\newblock Detection of biotic or abiotic stress in vineyards using thermal and rgb images captured via iot sensors.
\newblock \emph{IEEE Access}, 2023.

\bibitem[Hollmann et~al.(2023)Hollmann, M{\"u}ller, Eggensperger, and Hutter]{hollmann2023tabpfn}
Noah Hollmann, Samuel M{\"u}ller, Katharina Eggensperger, and Frank Hutter.
\newblock Tab{PFN}: A transformer that solves small tabular classification problems in a second.
\newblock In \emph{The Eleventh International Conference on Learning Representations}, 2023.
\newblock URL \url{https://openreview.net/forum?id=cp5PvcI6w8_}.

\bibitem[Kanaley et~al.(2023)Kanaley, Combs, Paul, Jiang, Bates, and Gold]{kanaley2023assessing}
Kathleen Kanaley, David~B Combs, Angela Paul, Yu~Jiang, Terry Bates, and Kaitlin~M Gold.
\newblock Assessing the capacity of high-resolution commercial satellite imagery for grapevine downy mildew detection and surveillance in new york state.
\newblock \emph{bioRxiv}, pp.\  2023--11, 2023.

\bibitem[Kerkech et~al.(2020)Kerkech, Hafiane, and Canals]{kerkech2020vine}
Mohamed Kerkech, Adel Hafiane, and Raphael Canals.
\newblock Vine disease detection in uav multispectral images using optimized image registration and deep learning segmentation approach.
\newblock \emph{Computers and Electronics in Agriculture}, 174:\penalty0 105446, 2020.

\bibitem[Liu et~al.(2019)Liu, Zhang, Chen, and Howell]{liu2019vineyard}
Di~Liu, Pangzhen Zhang, Deli Chen, and Kate Howell.
\newblock From the vineyard to the winery: how microbial ecology drives regional distinctiveness of wine.
\newblock \emph{Frontiers in Microbiology}, 10:\penalty0 2679, 2019.

\bibitem[McElfresh et~al.(2023)McElfresh, Khandagale, Valverde, Ramakrishnan, Goldblum, White, et~al.]{mcelfresh2023neural}
Duncan McElfresh, Sujay Khandagale, Jonathan Valverde, Ganesh Ramakrishnan, Micah Goldblum, Colin White, et~al.
\newblock When do neural nets outperform boosted trees on tabular data?
\newblock \emph{arXiv preprint arXiv:2305.02997}, 2023.

\bibitem[M{\"u}ller et~al.(2021)M{\"u}ller, Hollmann, Arango, Grabocka, and Hutter]{muller2021transformers}
Samuel M{\"u}ller, Noah Hollmann, Sebastian~Pineda Arango, Josif Grabocka, and Frank Hutter.
\newblock Transformers can do bayesian inference.
\newblock \emph{arXiv preprint arXiv:2112.10510}, 2021.

\bibitem[Picard \& Ahmed(2024)Picard and Ahmed]{picard2024fast}
Cyril Picard and Faez Ahmed.
\newblock Fast and accurate zero-training classification for tabular engineering data.
\newblock \emph{arXiv preprint arXiv:2401.06948}, 2024.

\bibitem[Vaswani et~al.(2017)Vaswani, Shazeer, Parmar, Uszkoreit, Jones, Gomez, Kaiser, and Polosukhin]{vaswani2017attention}
Ashish Vaswani, Noam Shazeer, Niki Parmar, Jakob Uszkoreit, Llion Jones, Aidan~N Gomez, {\L}ukasz Kaiser, and Illia Polosukhin.
\newblock Attention is all you need.
\newblock \emph{Advances in neural information processing systems}, 30, 2017.

\bibitem[Zhang et~al.(2019)Zhang, Huang, Pu, Gonzalez-Moreno, Yuan, Wu, and Huang]{zhang2019monitoring}
Jingcheng Zhang, Yanbo Huang, Ruiliang Pu, Pablo Gonzalez-Moreno, Lin Yuan, Kaihua Wu, and Wenjiang Huang.
\newblock Monitoring plant diseases and pests through remote sensing technology: A review.
\newblock \emph{Computers and Electronics in Agriculture}, 165:\penalty0 104943, 2019.

\bibitem[Zhao \& Efremova(2023)Zhao and Efremova]{zhao2023soil}
W.~Zhao and N.~Efremova.
\newblock Soil organic carbon estimation from climate-related features with graph neural network.
\newblock \emph{arXiv preprint arXiv:2311.15979}, 2023.

\end{thebibliography}
\bibliographystyle{iclr2024_conference}




 


\end{document}